\documentclass[twocolumn]{aastex63}

\usepackage{multirow}
\usepackage{booktabs}
\usepackage{graphicx}
\usepackage{amsmath}
\usepackage{longtable}
\usepackage{rotating}
\usepackage{enumerate}
\usepackage{color}

\newcommand{\uat}[2]{\href{http://vocabs.ands.org.au/repository/api/lda/aas/the-unified-astronomy-thesaurus/current/resource.html?uri=http://astrothesaurus.org/uat/#1}{#2 (#1)}}

\begin{document}

\title{The Appearance of Vortices in Protoplanetary Disks in Near-Infrared Scattered Light}

\author{Metea Marr}
\affiliation{Department of Physics, Simon Fraser University, Burnaby, BC, V5A 1S6, Canada}
\affiliation{Department of Physics \& Astronomy, University of Victoria, Victoria, BC, V8P 5C2, Canada}

\author{Ruobing Dong}
\affiliation{Department of Physics \& Astronomy, University of Victoria, Victoria, BC, V8P 5C2, Canada}
\email{rbdong@uvic.ca}

\begin{abstract}

Azimuthally asymmetric structures have been discovered in millimeter continuum emission from many protoplanetary disks. One hypothesis is that they are vortices produced 
by the Rossby wave instability, for example at edges of planet-opened gaps or deadzones.
Confirming the vortex nature of these structures will have profound implications to
planet formation. One way to test the hypothesis is to compare the observed morphology of vortex candidates in near-infrared scattered light with theoretical expectations.
To this end, we synthesize the appearance of vortices in $H$-band polarized light
by combining hydrodynamic and radiative transfer simulations of the Rossby wave instability at a deadzone edge.
In a disk at 140 pc, at the peak in its evolution a vortex at 65 au may appear as a radially narrow arc $50\%-70\%$ brighter compared with an axisymmetric disk model. The contrast depends on the inclination of the disk and the position angle of the vortex only weakly. 
Such contrast levels are well detectable in imaging observations of bright disks using instruments such as VLT/SPHERE, Subaru/SCExAO, and Gemini/GPI. A vortex also casts a shadow in the outer disk, which may aid its identification. Finally, at modest to high inclinations (e.g., $60^\circ$) a vortex may mimic a one-armed spiral. In the HD 34282 disk, such a one-armed spiral with a shadowed region on the outside has been found in scattered light. This feature roughly coincides with an azimuthal asymmetry in mm continuum emission, signifying the presence of a vortex.

\end{abstract}



\section{Introduction}\label{sec:intro}

Imaging observations with sufficiently high angular resolution and sensitivity have revealed structures in a number of protoplanetary disks. Such observations are typically carried out in two spectral windows. At near infrared (NIR) wavelengths, observations probe starlight being scattered by small dust grains at the disk surface, typically $\micron$-sized or smaller. At (sub-)millimeter (mm) wavelengths, radio interferometers such as the Atacama Large millimeter Array (ALMA) image line and continuum emissions from gas and dust, with the latter being typically sub-mm in size. Millimeter continuum observations have revealed structures in many disks \citep{andrews20}. Some of these disks have also been imaged in NIR scattered light to enable multi-wavelength studies of the same structures, yielding insights into their origins \citep[e.g.,][]{dong17j1604}.

Many disks show concentric rings in mm continuum emission \citep[e.g.,][]{andrews18, long18,  vandermarel19, francis20}. In some of them, the rings host large scale azimuthal asymmetries \citep[e.g.,][]{isella13, perez14, vandermarel13, casassus13} at stellocentric distances of $\sim$10 to $\sim$100 au \citep{vandermarel21}. One possible explanation for these asymmetries is that they are dust traps produced by vortices triggered by the Rossby wave instability \citep[RWI;][]{li00, li01}. The RWI may be excited at steep density transitions, such as the edge of planet-opened gaps \citep{hammer17, hammer19, hallam20}, or at the edge of viscosity transitions \citep[e.g., deadzone edges,][]{regaly12, flock15}. Once formed, a vortex may trap dust of certain sizes inside \citep{birnstiel13}. Multi-fluid simulations with both gas and dust have shown that vortices may appear as emission clumps at mm wavelengths similar to the observed azimuthal asymmetries \citep{zhu14stone, baruteau19}. However, definitive evidence is currently lacking. Other mechanisms have been proposed to explain azimuthal symmetries as well. For example, they may be horseshoes at the edges of eccentric cavities harboring massive companions \citep[e.g.,][]{ragusa17, calcino19}.

The origin of observed asymmetries is a key to the study of planet formation. If they are indeed dust trapping vortices, the enrichment of dust in them may facilitate planetesimals formation via, e.g., the streaming instability \citep[e.g.,][]{youdin07si, bai10si, li19si}. In addition, if they are produced at the edges of planet-opened gaps, they may be the signposts of planets.

How do we test the hypothesis that observed asymmetries are vortices? One avenue may be to search for the anticyclonic gas motions inside these structures using gas observations \citep{huang18vortex, robert20}. This technique is in principle straightforward; however, the required high spatial resolution and sensitivity are challenging, and the applications to real systems are inconclusive \citep[e.g.,][]{boehler21}. Vortices may also have distinct spectral indices at mm wavelengths due to differential trapping of dust of different sizes \citep{birnstiel13}. However, new explorations in dust scattering at mm wavelengths necessitate the need to revisit the models \citep{liu19, zhu19scattering}.

Another way to test the vortex hypothesis is to compare the morphology of a vortex candidate at multiple wavelengths with the theoretically expected appearance of a vortex at the corresponding wavelengths --- a true vortex should look like a vortex at any wavelengths. To do so, we need to understand how vortices appear in various observations. While the morphology of dust clumps produced by vortices at mm wavelengths has been well studied \citep[e.g.,][]{zhu14stone, baruteau16}, the morphology of vortices in NIR scattered light lacks a thorough understanding. 

In this work, we study synthetic observations of vortices triggered by the RWI at the deadzone edge in NIR scattered light using hydrodynamics and radiative transfer simulations. We introduce our numerical setups in \S~\ref{sec:simulations}, present the results in \S~\ref{sec:results}, and summarize and discuss our findings in \S~\ref{sec:discussions}.


\section{Simulations}\label{sec:simulations}

We use FARGO3D \citep{benitezllambay16} hydrodynamic simulations to model a locally isothermal and non-self-gravitating protoplanetary disk subject to the RWI and vortex formation at a deadzone edge. We follow \citet{huang19} in setting up the simulations. We then feed the resulting disk structures into HOCHUNK3D \citep{whitney13} radiative transfer simulations to visualize the disk in NIR polarized scattered light. We follow \citet{dong15gap} in joining hydro and radiative transfer simulations. 

\subsection{Hydrodynamic Simulations}\label{sec:hydro}

Our grid is two dimensional (2D) with size (2048, 3072) in the radial $R$ and azimuthal $\phi$ directions, respectively, and with uniform spacing in both directions. We use a full disk with $\phi$ ranging from $0$ to $2\pi$ and $R$ ranging from $0.2R_0$ to $4.5R_0$, where $R_0$ is the code length unit. We use symmetric radial boundary conditions. The disk scale height $H$ as a function of $R$ is given by:
\begin{equation}\label{eq:scaleheight}
    \frac{H(R)}{R} = 0.05 \left(\frac{R}{R_0}\right)^{1/4}.
\end{equation}
The initial gas surface density profile $\Sigma(R)$ is given by:
\begin{equation}
    \Sigma(R) = \Sigma_0 \left(\frac{R}{R_0} \right)^{-1}
    \label{eq:sigma}
\end{equation}
where $\Sigma_0$ is a constant to be normalized by the total disk mass. The viscosity is characterized by the \citet{shakura73} $\alpha$ parameter. A deadzone model is implemented by varying $\alpha$ with $R$ as:
\begin{equation}
    \alpha(R) = \alpha_0 - \frac{\alpha_0 - \alpha_{DZ}}{2}\left[1 - \tanh{\left(\frac{R - R_{DZ}}{\Delta_{DZ}}\right)}\right],
\end{equation}
where the viscosity inside the deadzone $\alpha_{DZ} = 10^{-5}$, the viscosity outside the deadzone $\alpha_0 = 10^{-3}$, and the transition occurs at $R_{DZ} = 1.5R_0$.

Following \citet{huang19}, we experiment with different radial widths of the viscosity transition region $\Delta_{DZ}$ to search for vortices with the largest surface density contrasts at the peak. Such vortices are expected to be the most prominent and the easiest to detect in NIR scattered light. In total three models are tested:
\begin{itemize}
  \item Model Scale Height (SH), in which $\Delta_{DZ}=H$ is the local disk scale height.
  \item Model Half-Scale Height (H-SH), in which $\Delta_{DZ}=H/2$.
  \item Model Quarter-Scale Height (Q-SH), in which $\Delta_{DZ}=H/4$.
\end{itemize}
The hydrodynamic simulations are run for 3000 orbits at $R_0$. 

\subsection{Radiative Transfer Simulations}\label{sec:rt}

We puff up 2D surface density maps of the disk from hydro models in the vertical direction to restore its 3D geometry and to simulate NIR scattered light observations. This is done using a Gaussian vertical density profile, i.e., assuming the disk is vertically isothermal:
\begin{equation}\label{eq:rho}
    \rho(R,z) = \rho_0(R)e^{-\frac{z^2}{2[H(R)]^2}}
\end{equation}
where $z$ is the vertical height from the midplane and $\rho_0(R)$ is the density at $z=0$ normalized by $\Sigma(R)$. 

Our assumption that the vertical density distribution is the same inside and outside a vortex is justified. Both analytical and numerical studies have shown that the RWI and the resulting vortex formation process are 2D in nature \citep{meheut10, meheut12, lin12rwi, richard13, lin14}. In our locally isothermal non-self-gravitating disks, the differences in the gas surface density and vorticity in vortices between 2D and 3D simulations are negligible \citep{zhu14votices}. In addition, inside a vortex, vertical gas motions and density stratification (i.e., deviations from Eqn.~\ref{eq:rho}) are almost absent (\citealt{lin12rwi}, \citealt{lin18vortex}). Vortices may develop internal structures in non-isothermal disks or when disk self-gravity becomes important \citep{meheut12, lin12, lin18vortex}.

The 3D disk has 1126, 384, 122 cells in spherical coordinates ($R$, $\theta$, $\phi$) ($\theta$ is the polar angle). In the polar direction the grid extends to $\pm25^\circ$ from the disk midplane, or $\sim$8 scale heights at the vortex location. We bin every 2 radial cells and 8 azimuthal cells in hydro models to one cell in radiative transfer calculations. Convergence tests have shown that the resulting spatial resolution is sufficient to resolve the vortices in all dimensions.
We set $R_0=50$ au, which places the vortex at $\sim$65 au, and normalize the initial disk mass to 0.03 solar masses within 225 au.
In HOCHUNK3D we pad an axisymmetric inner disk between the dust sublimation radius and the hydro inner boundary ($0.2R_0$) with 100 radial cells by extrapolating the azimuthally averaged hydro disk at its inner edge inward assuming the same radial surface density profile as in Eqn.~\ref{eq:sigma}.

Our hydro simulations are gas only, while scattered light is determined by the distribution of small dust typically sub-$\micron$ in sizes. Such dust usually has Stokes numbers much smaller than unity, and is expected to be well-coupled with the gas. We thus assume a constant dust-to-gas mass ratio, 1:100, throughout the disk. We note that if grain growth and evolution occur inside vortices, dust size distributions may be modified \citep{li20}. We assume interstellar medium dust \citep{kim94} with a power-law size ($s$) distribution $n(s)\propto s^{-3.5}$ between $s = 0.002 - 0.25 \micron$. The optical properties of the dust can be found in Fig. 2 in \citet{dong12cavity}.

We assume a star with a radius of 2.4 $R_\odot$ and a temperature of 4400 K in radiative transfer simulations. We produce $H$-band ($1.6\mu$m) polarized intensity (PI) images at inclinations $i$ of 0$^\circ$, 30$^\circ$, and 60$^\circ$, with the vortex placed on the major axis (position angle PA$_{\rm vortex}$ = $90^\circ$), minor axis (PA$_{\rm vortex}$ = $0^\circ$, far side; and $180^\circ$, near side), and in between positions (PA$_{\rm vortex}$ = $45^\circ$ and $135^\circ$). Note that since we are viewing a non-axisymmetric structure at a cone-shaped surface (because the disk is optically thick at NIR wavelengths), the structure appears differently at these position angles when $i\neq0$ \citep{dong16armviewing}. 

Synthetic images from radiative transfer simulations are at ``full resolution''. To mimic real observing conditions, we convolve images by a Gaussian point-spread function (PSF) to achieve an angular resolution of 0.04\arcsec\ (the diffraction limited angular resolution of 8-meter telescopes) assuming the disk is at 140 pc. Images are produced using 4 billion photon packets, and will be shown in linear stretch. Tests show that the noises introduced by this finite number of photon packets is at the percent level in convolved images.


\section{Results}\label{sec:results}

\begin{figure}
\centering
\includegraphics[width=0.45\textwidth,angle=0]{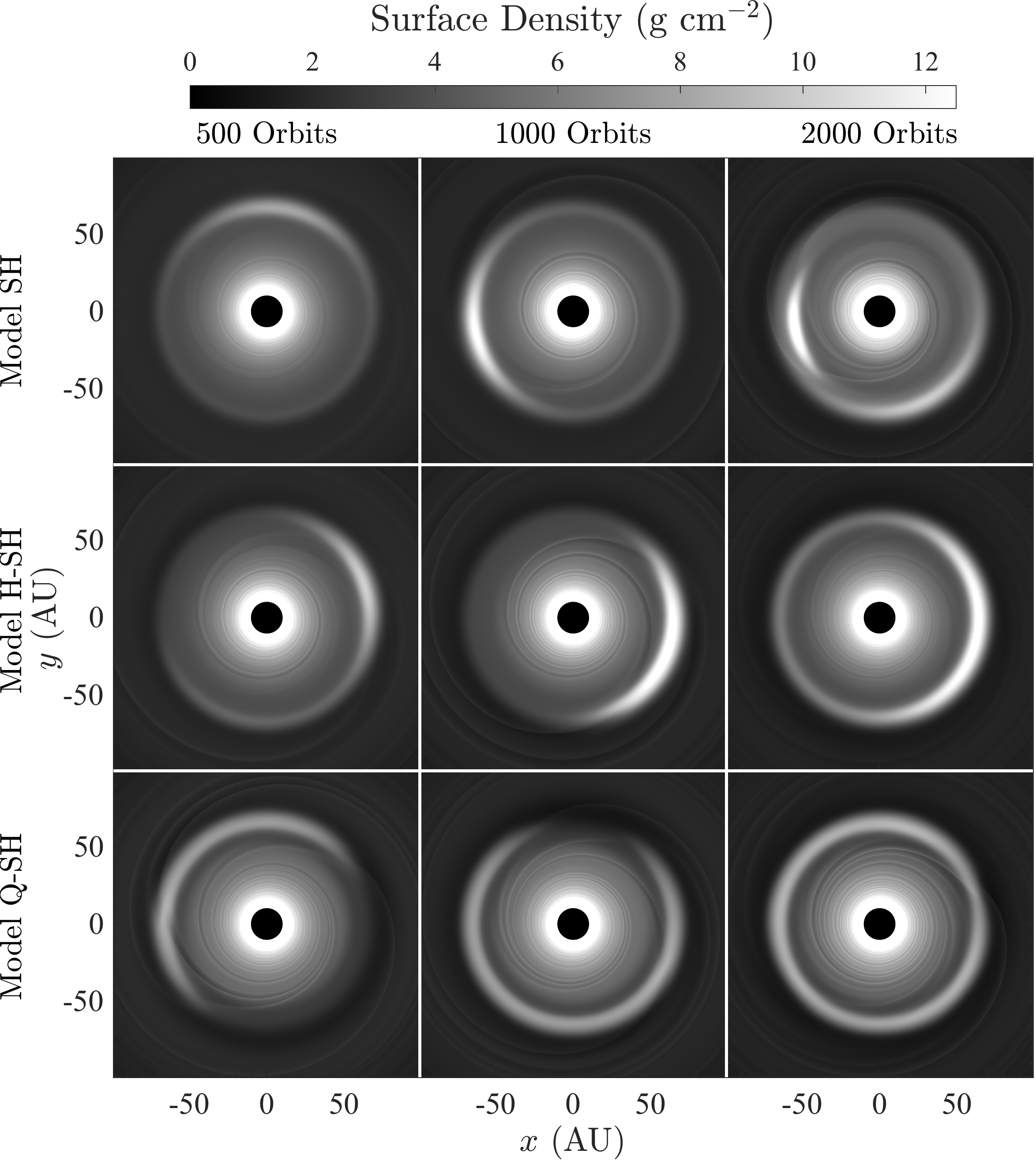}
\includegraphics[width=0.02\textwidth,angle=0]{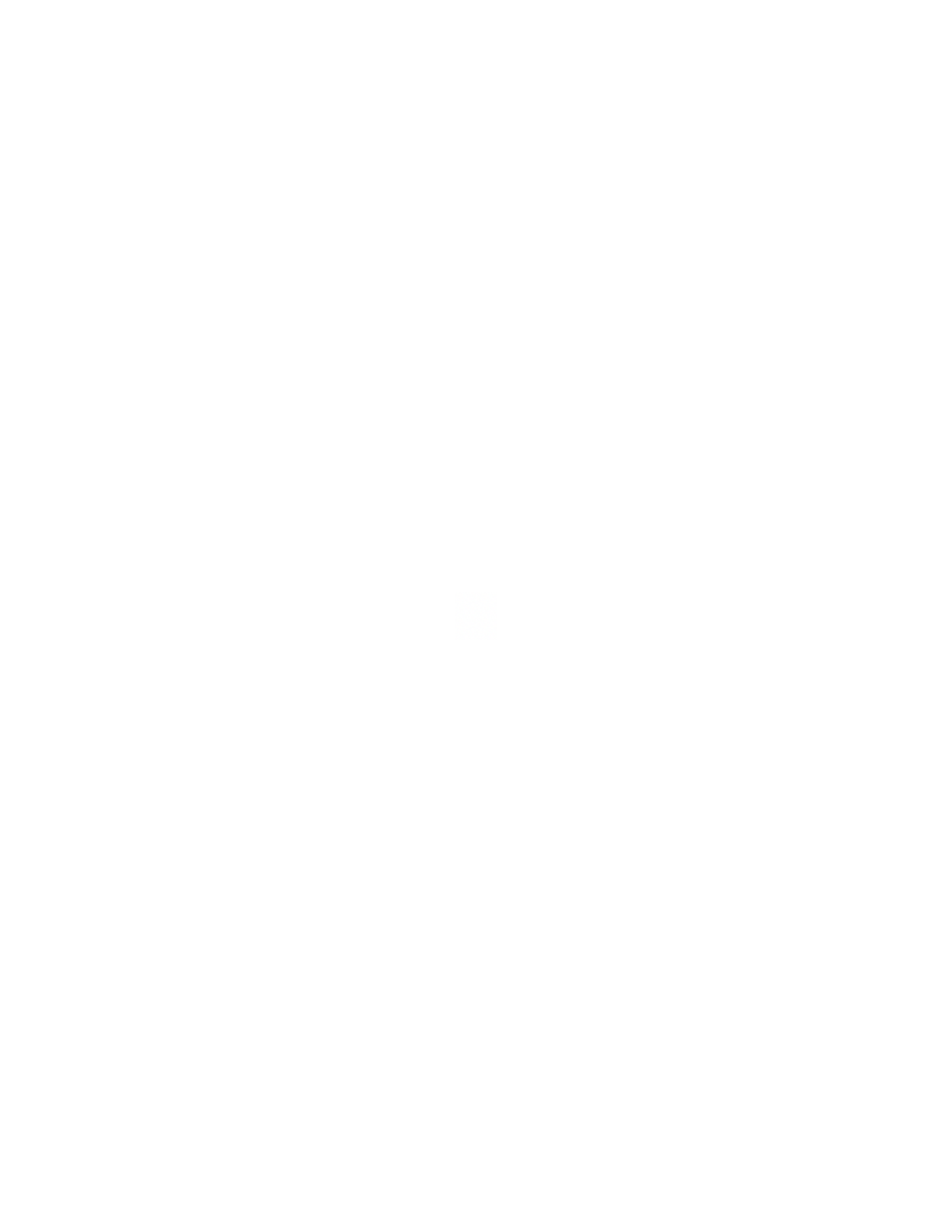}
\includegraphics[width=0.45\textwidth,angle=0]{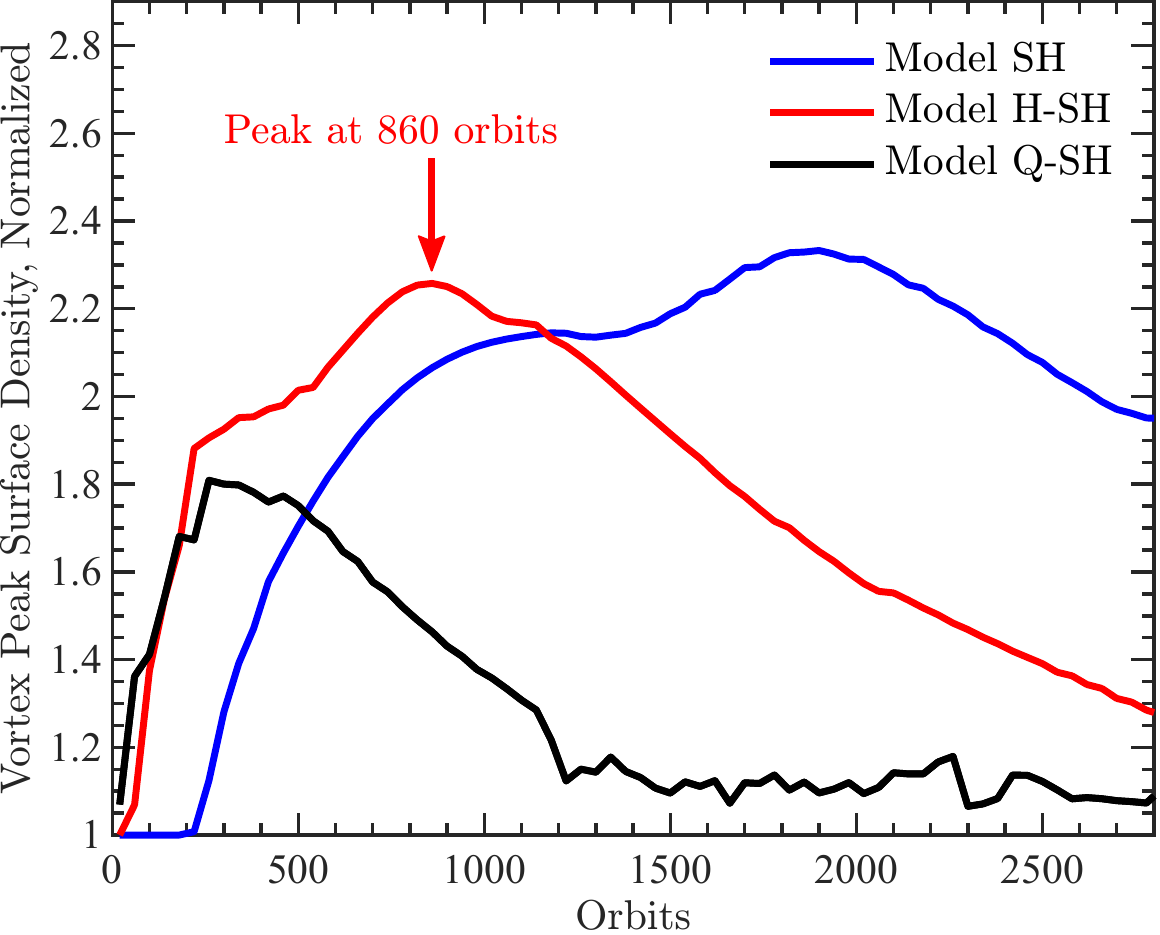}
\centering
\caption{Top: Surface density maps for the three models at 500, 1000, and 2000 orbits. The region inside the hydro inner boundary is masked out. Bottom: Temporal evolution of the vortex amplitudes, showing the surface density at the peak of the primary vortex normalized by the azimuthally averaged surface density at that radius. The primary vortex in Model H-SH peaks at 860 orbits. 
The FITs files for the top panel are available as Astrophysical Journal online supplemental material.
}
\label{fig:sigma}
\end{figure}

Fig.~\ref{fig:sigma} shows the surface density maps of the models at 500, 1000, and 2000 orbits. In all cases, the RWI is excited around the viscosity transition region. Initially a number of small vortices form, which quickly merge to form a single big vortex, before it is gradually elongated in the azimuthal direction. As we are interested in finding the most prominent vortex, we show the temporal evolution of the maximum surface density in the merged vortex relative to the azimuthally averaged background at the bottom. The vortices in Models SH and H-SH reach roughly the same peak amplitudes, $\sim$2.3$\times$ the azimuthally averaged surface density at their radii. The vortex in Q-SH is significantly weaker. Between SH and H-SH, we choose H-SH for further investigation in scattered light because when the primary vortex in Model SH peaks at $\sim1800$ orbits, a second generation vortex has emerged (visible in the 2000 orbits panel in Fig.~\ref{fig:sigma}), which complicates the characterization of the primary vortex. 

\begin{figure*}
\includegraphics[width=\textwidth,angle=0]{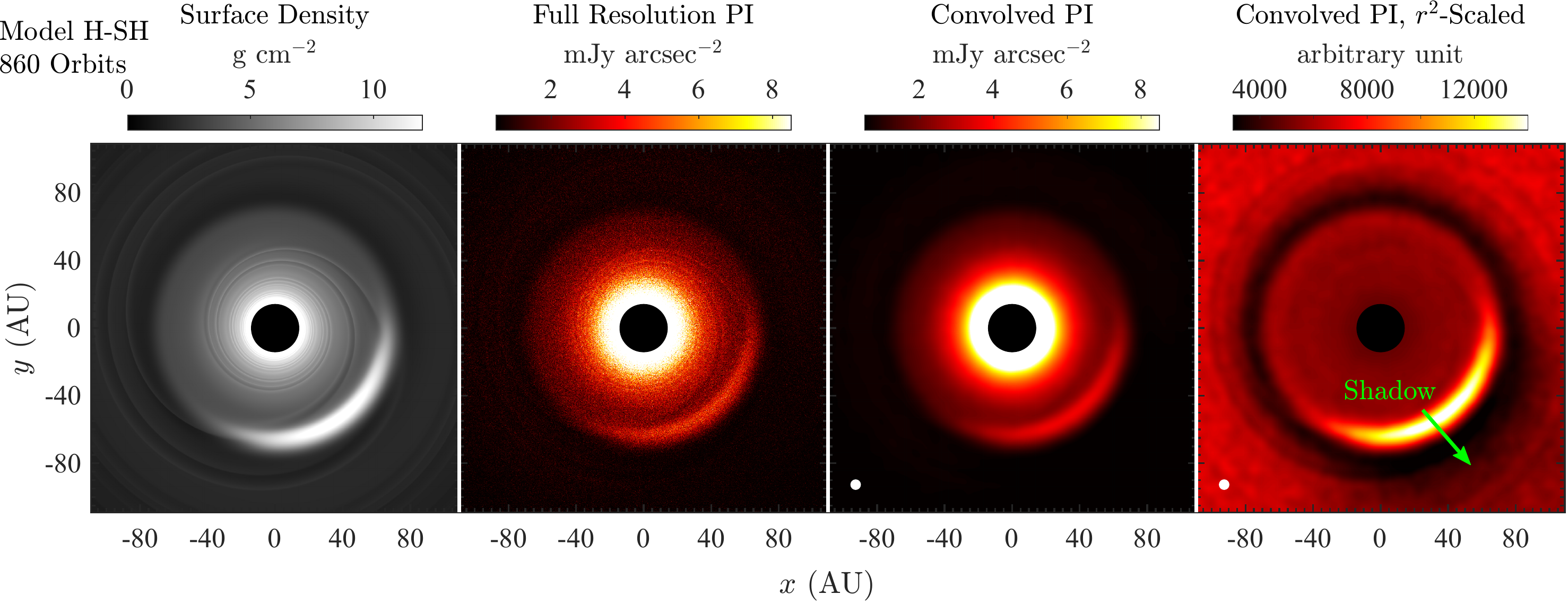}
\caption{The surface density map (left), the full resolution $H$-band polarized intensity (PI) image (middle-left), the convolved image (middle-right), and the convolved image scaled by $r^2$ ($r$ is the stellocentric distance) for Model H-SH at 860 orbits, when the vortex peaks. In the convolved images the disk is assumed to be at 140 pc and the PSF size is 0.04\arcsec (marked at the lower left corner). In all panels the inner $0.1\arcsec$ (14 AU) in radius is masked out to mimic the effect of an inner working angle typically achieved in today's observations. The vortex is clearly visible in the synthetic images. The shadow cast by the vortex in the outer disk is better seen in the $r^2$-scaled convolved image. 
The FITs files are available as Astrophysical Journal online supplemental material.
}
\label{fig:image}
\end{figure*}

\begin{figure}
\includegraphics[width=0.45\textwidth,angle=0]{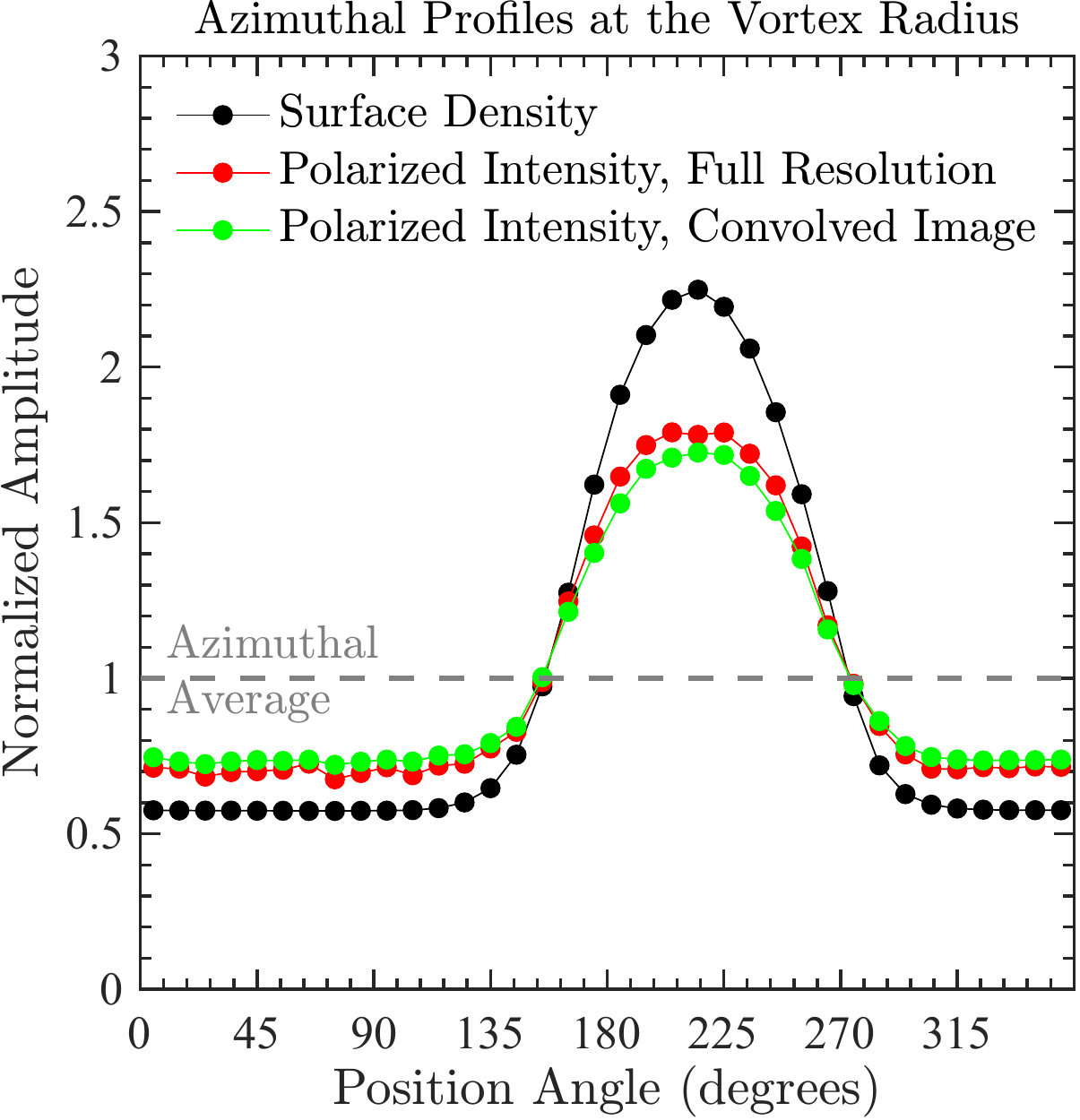}
\caption{The azimuthal profiles of Model H-SH at 860 orbits in surface density, full resolution $H$-band polarized intensity, and convolved $H$-band polarized intensity (i.e., the first three panels in Fig.~\ref{fig:image}). The measurements are taken at the radius of the vortex peak and averaged over radii of 4 au across that radius (roughly the local disk scale height). The amplitudes are normalized by the azimuthal averages (gray dashed line at 1). In the convolved image the vortex peak is $\sim$70\% brighter than the azimuthally averaged background, and $\sim$140\% brighter than the faint region on the opposite side.}
\label{fig:azimuthal}
\end{figure}

The vortex in Model H-SH peaks at 860 orbits (Fig.~\ref{fig:sigma}). Its surface density map at that epoch and the corresponding synthetic scattered light images at face-on are shown in Fig.~\ref{fig:image}. The azimuthal profiles at the vortex radius, 67 au in surface density and 62 au in scattered light, are shown in Fig.~\ref{fig:azimuthal}. Overall, the vortex appears as a radially narrow arc.
The radial full width half maximum of the vortex relative to the background is $\sim$11 au, or $\sim$3 local disk scale heights. This is about half the PSF size (0.04\arcsec, or 5.6 au at 140 pc). Therefore, the vortex in the convolved image is marginally resolved, and the peak contrasts in the full resolution and convolved images are nearly the same.
The vortex peak in the convolved image is $\sim$70\% brighter than the azimuthally averaged background, and $\sim$140\% brighter than the region on the opposite side at the same radius.
NIR polarized light imaging observations today using instruments such as VLT/SPHERE, Subaru/SCExAO, and Gemini/GPI can reach a precision level of $<$10\% in local surface brightness registration for bright disks (\citealt{pinilla18}, Fig. 4; \citealt{muroarena20}, Fig. 2; and reach percent level precision in azimuthally averaged surface brightness, \citealt{ren21}). The vortex in our model is expected to be easily detectable.

\begin{figure*}
\centering
\includegraphics[width=0.9\textwidth,angle=0]{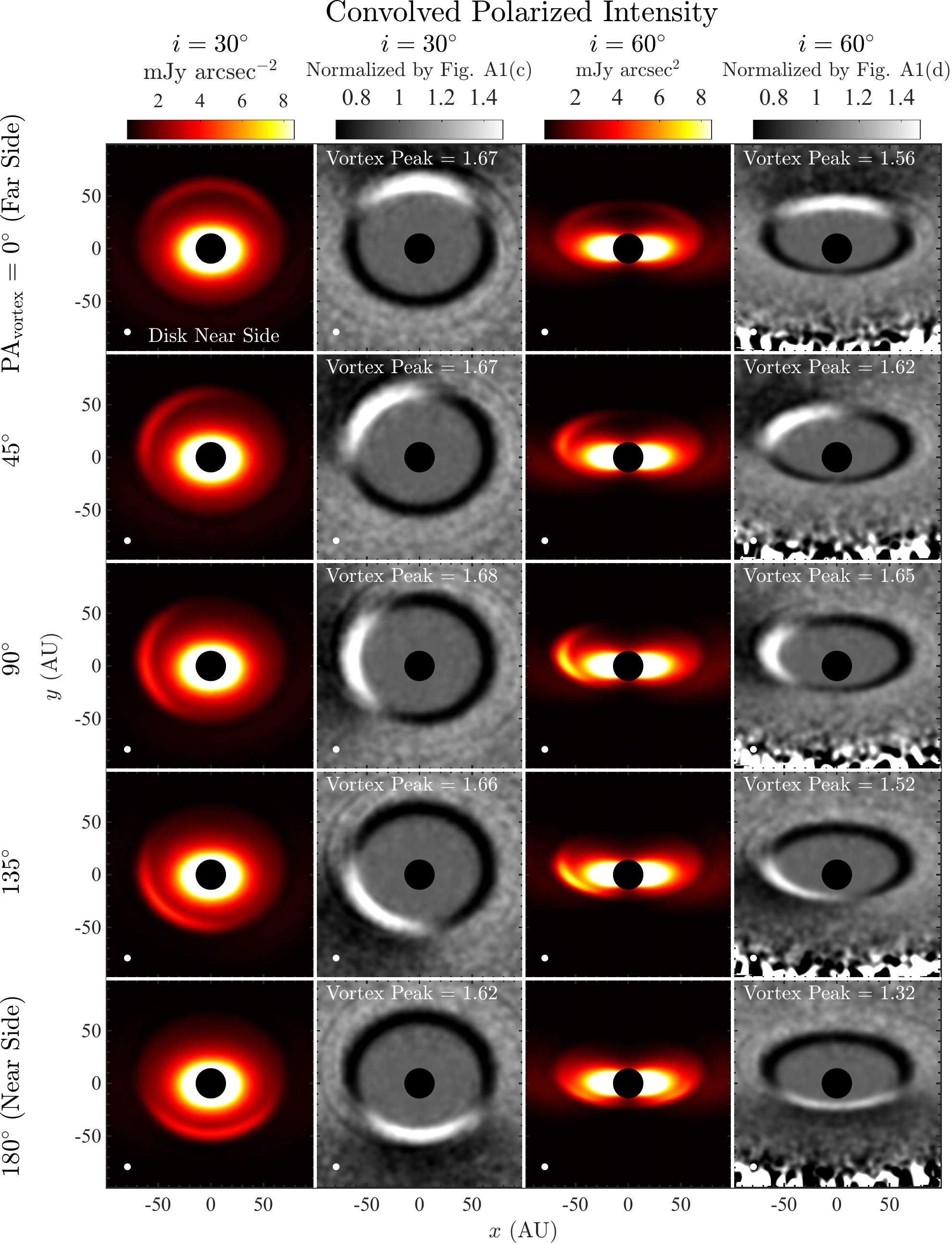}
\caption{The red-hot columns ($1^{\rm st}$ and $3^{\rm rd}$) show convolved synthetic $H$-band polarized light images of Model H-SH at 860 orbits at two inclinations ($i=30^\circ$ and $60^\circ$) and with the vortex at five position angles (top to bottom: PA$_{\rm vortex}=0^\circ$, $45^\circ$, $90^\circ$, $135^\circ$, and $180^\circ$). The major axis of the disk (a.k.a. the position angle of the disk) is in the horizontal direction, and it is inclined such that the south side is the near side \citep[see Fig. 1 in][for the definitions of ``near'' and ``far'' sides]{dong16armviewing}. The object is assumed to be at 140 pc and the PSF size is 0.04\arcsec (marked at the lower left corner). 
The gray columns ($2^{\rm nd}$ and $4^{\rm th}$) show convolved model images normalized by convolved images of asymmetric disk models (Fig.~\ref{fig:azi_ave}). 
These normalized images are produced in order to remove the intrinsic azimuthal variations in disk images caused by inclinations.
The peak of the vortex in these normalized images is labeled on the top of each panel.  In all panels the inner $0.1\arcsec$ (14 AU) in radius is masked out to mimic the effect of an inner working angle typically achieved in today's observations. The vortex manifests itself as a bright arc, with amplitudes reaching $\sim1.5-1.7$ in the normalized images at most viewing angles. See \S\ref{sec:results} for details. 
The FITs files are available as Astrophysical Journal online supplemental material.
}
\label{fig:ipa}
\end{figure*}

\begin{figure}
\centering
\includegraphics[width=0.5\textwidth,angle=0]{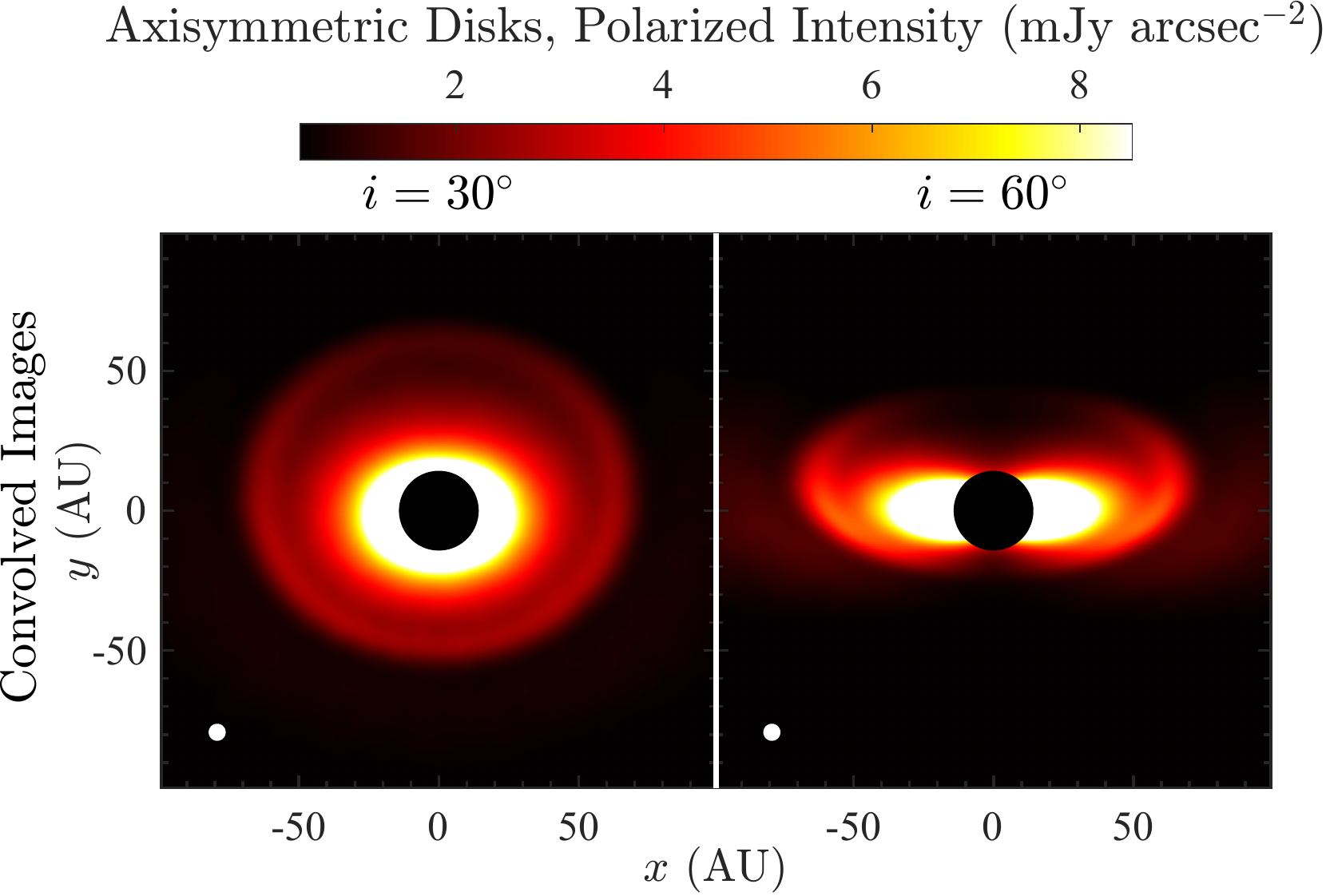}
\caption{Convolved $H$-band polarized intensity images for Model H-SH at 860 orbits at $i=30^\circ$ and $60^\circ$, but with the surface density azimuthally averaged. In other words, the surface density in the disk is axisymmetric, while it has the same radial profile as that in Model H-SH. The model images have a ring at the radial location of the vortex in Model H-SH. The rings have azimuthal variations in scattered light due to finite inclinations. 
The FITs files are available as Astrophysical Journal online supplemental material.
}
\label{fig:azi_ave}
\end{figure}

Fig.~\ref{fig:ipa} shows the convolved $H$-band images of Model H-SH at 860 orbits (the ``red-hot'' panels), as well as the convolved images normalized by axisymmetric model images (the gray panels). In total we show 10 viewing geometries: two inclinations ($i=30^\circ$ and $60^\circ$), and 5 position angles for the vortex (PA$_{\rm vortex}=0^\circ$, $45^\circ$, $90^\circ$, $135^\circ$, $180^\circ$).
In scattered light, even an axisymmetric structure in density distribution (e.g., a ring) displays azimuthal variations at $i\neq0$, mainly caused by the angular dependence of scattering \citep[e.g., Fig. 1 in][]{dong17gap}.
We do not want such inclination-based intensity variations to be confused with intensity variations caused by a vortex.
To define the contrast of the vortex at $i\neq0$,
we normalize the convolved images by that of a disk with an axisymmetric surface density distribution at the same inclination.
The latter disk has the same radial surface density profile as in the vortex disk model; in other words it is the azimuthally averaged version of the model. The synthetic images of these axisymmetric disk models are shown in Fig.~\ref{fig:azi_ave}.
The peak of the vortex in those ``normalized'' images, which we defined as ``contrast'', is labeled on the panels. 

Due to the confusion with inclination-based azimuthal variations, it can be difficult to recognize the vortex in the convolved images when it is on the disk minor axis (PA$_{\rm vortex}=0$ or 180$^\circ$) and at higher inclinations. When the vortex is away from the disk minor axis, the asymmetry with respect to the minor axis facilitates its identification.
The absolute surface brightness of the vortex depends on its position angle; for example, it is brighter when it is on the near side (PA$_{\rm vortex}=180^\circ$) due to the forward scattering of dust. In addition, it is easier to separate the vortex from the bright inner disk when the vortex is on the major axis due to projection effects. 
Meanwhile, the contrast of the vortex does not sensitively depend on its position angle. At $i=30^\circ$ the vortex is always $\sim$60\%$-70$\% brighter than the same region in the axisymmetric model; at $i=60^\circ$ the vortex is slightly less prominent, being $\sim$50\%$-65$\% brighter at all PA$_{\rm vortex}$ except at $180^\circ$ (near side), where it is only $\sim$30\% brighter. Such contrasts are well detectable in scattered light imaging observations of bright disks.

There are two features in our synthetic images that are worth highlighting. First, the vortex casts a shadow in the outer disk. The vortex is brighter than the surrounding because it has a higher surface (defining as the optical depth $\tau=1$ surface from the star at the observing wavelength), thus better illuminated (\citealt{takami14}). While the outer disk beyond the radius of the vortex is faint overall as the region is shadowed by the ring of material at the deadzone transition region \citep{ueda19}, the vortex produces an enhanced shadow in the outer disk at its position angle. This is best seen at face-on, when the outer disk is free from inclination-induced azimuthal variations (right panel, Fig.~\ref{fig:image}). Meanwhile it is also visible in inclined disks (see the normalized convolved images in Fig.~\ref{fig:ipa}). The presence of this shadow may provide additional evidence for the vortex. Similar shadow features in the outer disk caused by locally elevated disk surface have been seen in the case of circumplanetary disks \citep{weber21}.

\begin{figure}
\includegraphics[width=0.5\textwidth,angle=0]{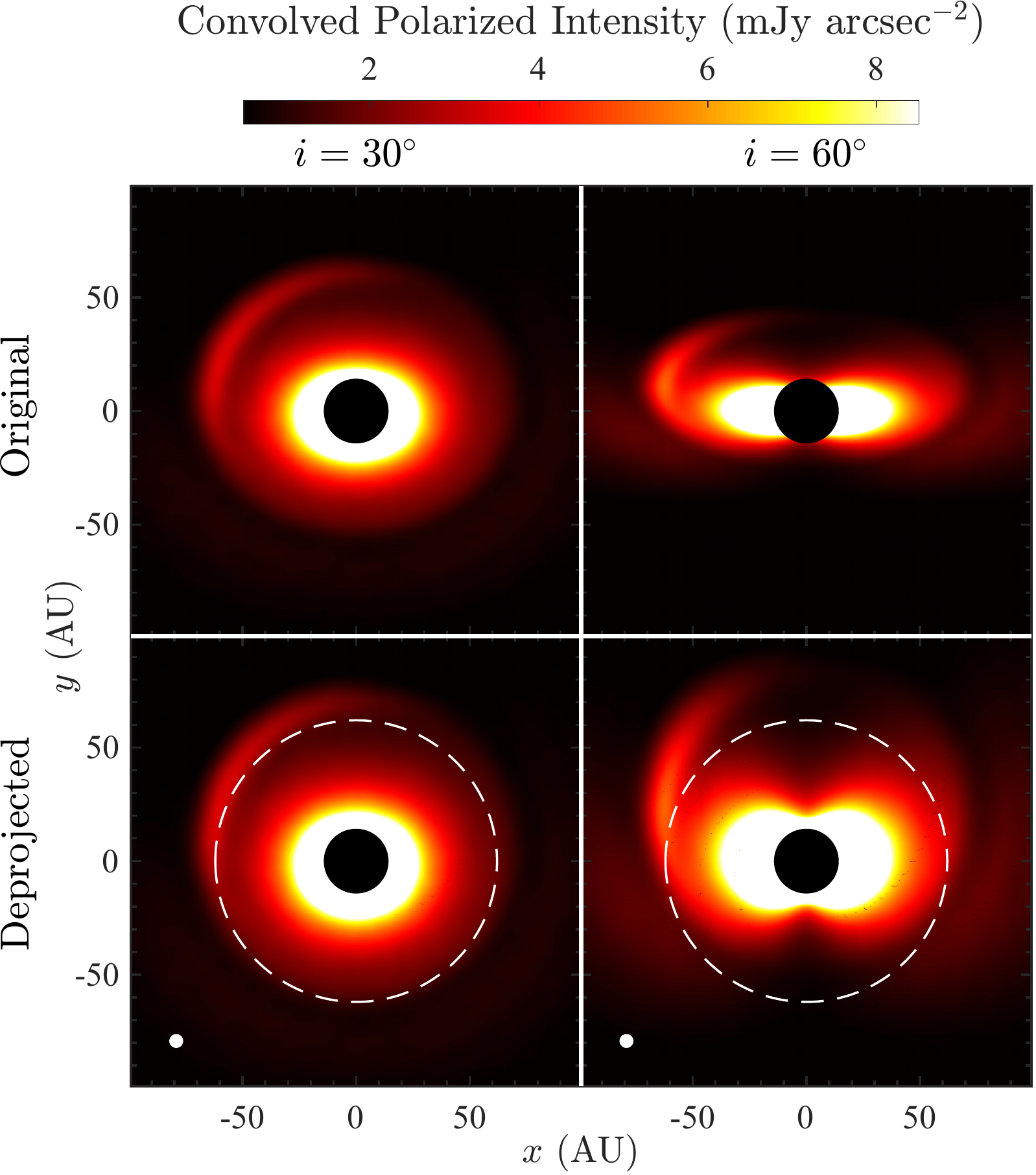}
\caption{Convolved $H$-band polarized intensity images for Model H-SH at 860 orbits at $i=30^\circ$ and $60^\circ$, with the vortex at the position angle of $45^\circ$ (the second row in Fig.~\ref{fig:ipa}). The top row shows the original images. The bottom row shows deprojected images (a linear stretch along the minor axis with a factor of $1/\cos{i}$), produced using {\tt diskmap} \citep{stolker16}. The dashed circle in the deprojected images marks the location of the ring on which the vortex is located in the face-on view (e.g., Fig.~\ref{fig:image}). A simple deprojection does not restore the true face-on view of the disk, particularly at high inclinations.}
\label{fig:deproj}
\end{figure}

Secondly, at $i\neq0$ the appearance of the vortex may be confused with a one-armed spiral when it is not on the minor axis. In principle, the vortex arc is part of a circle with zero pitch angle (the angle between the elongation of the arc and the azimuth). In inclined disks, however, axisymmetric density features may not center on the star in scattered light \citep{ginski16}. Thus it is difficult to tell whether the pitch angle of an arc-like feature is zero or not. In general the ``true'' face-on view of a disk in scattered light cannot be restored from an actual observation by performing a simple deprojection (Fig.~\ref{fig:deproj}; also see \citealt{dong16armviewing}), unless the shape of the disk surface in scattered light is well known \citep{stolker16}. A circular arc may thus be confused from a ``one-armed spiral'', particularly at high inclinations.

\begin{figure*}
\centering
\includegraphics[width=\textwidth,angle=0]{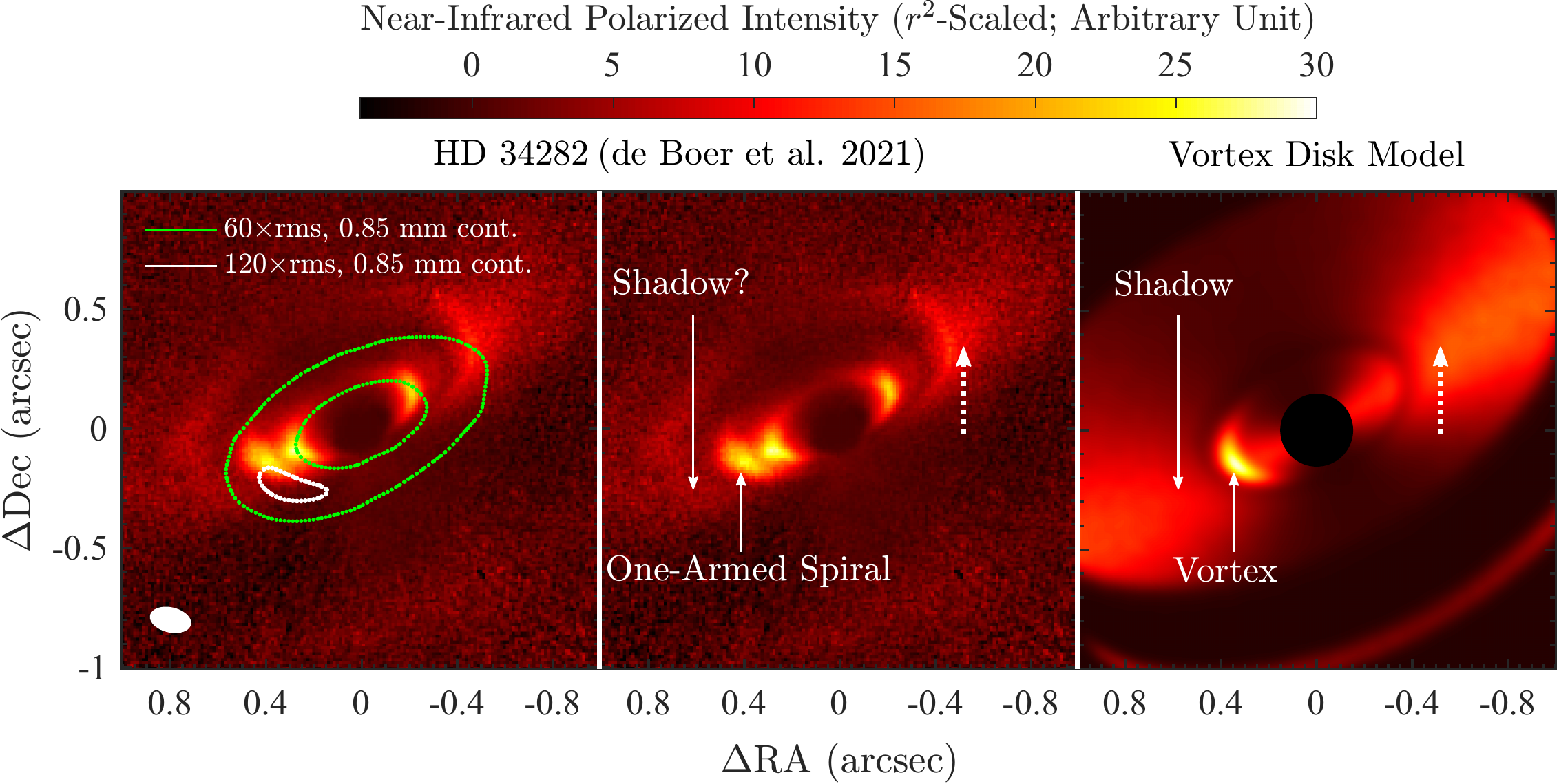}
\caption{Left: VLT/SPHERE $J$-band polarized intensity image of HD 34282 \citep{deboer21}. The 0.85 mm (351 GHz) ALMA Band 7 continuum emission with superuniform weighting (beam size $0.17\arcsec\times0.10\arcsec$; lower left corner) is shown
in green (white) contours at 60$\times$ (120$\times$) the root mean square (rms) noises. The green contours trace out a ring in the 0.85 mm emission, and the white contours trace out an azimuthal asymmetry. The ALMA data is from programme 2013.1.00658.S, originally presented by \citet{vanderplas17}. Middle: The same as the left panel, but without the ALMA contours. Right: Convolved image for Model H-SH at 860 orbits in NIR polarized scattered light. All scattered light images have been scaled by the square of the on-sky projected stellocentric distance. The model has been spatially rescaled and positioned to match the size and the viewing geometry of the HD 34282 disk ($i\sim56^\circ$ and disk position angle $\sim118^\circ$, \citealt{deboer21}). The vortex in the model appears as a one-armed spiral, resembling the observed one-armed spiral in the HD 34282 disk. Meanwhile the vortex in the model casts a shadow in the outer disk. In the HD 34282 disk the same is observed, i.e., the East side of the disk outside the one-arm spiral (pointed by the arrow ``Shadow'') is fainter than the West side on the opposite side (pointed by the dotted arrow). The ALMA contours show that the mm emission from the disk is ring like, with a major azimuthal asymmetry on the southeast side. The NIR one-armed spiral roughly coincides with the mm azimuthal asymmetry. 
The small difference between the two in location may be caused by the the difference between a surface feature (the former) and a midplane feature (the latter) in a projected view \citep[Fig. 4 in][]{dong18mwc758}.
The FITs file for the model is available as Astrophysical Journal online supplemental material.
}
\label{fig:hd34282}
\end{figure*}

The two features can both be seen in the HD~34282 disk. In Fig.~\ref{fig:hd34282} we compare the SPHERE NIR scattered light image of the disk with our model image at the same viewing angle. A one-armed spiral (feature B1 in \citealt{deboer21, quiroz21}) has been identified in the HD 34282 disk. The feature appears similar to the vortex in our model. In addition, the East side of the HD 34282 disk outside the one-arm spiral is fainter than the West side at the same radii. This feature matches well with the shadow cast by the vortex in the model. Excitingly, the HD 34282 disk is a ring disk in mm continuum emission, and the ring harbors an azimuthally asymmetric structure \citep{vanderplas17}, roughly coinciding with the one-armed spiral in scattered light (see the ALMA contours). 
Because mm continuum emission is expected to originate from a thin layer of dust at the disk midplane, the small difference between the two structures in location (visible in the left panel in Fig.~\ref{fig:hd34282}) may be caused by the the difference between a surface feature (the NIR one-armed spiral) and a midplane feature (the mm azimuthal asymmetry) in a projected view \citep[Fig. 4 in][]{dong18mwc758}.
These observations suggest that the feature may be a vortex.

\section{Summary and Discussions}\label{sec:discussions}

In this work we carry out 2D hydrodynamic simulations of protoplanetary disks, in which the Rossby wave instability is excited at a deadzone edge to form vortices (Fig.~\ref{fig:sigma}). We 
post-process the model with the strongest vortex in radiative transfer simulations to produce synthetic $H$-band polarized scattered light images. Our results show that imaging observations today using instruments such as VLT/SPHERE, Subaru/SCExAO, and Gemini/GPI are capable of detecting vortices in near-infrared scattered light.

At face-on, at its peak the vortex in the model H-SH appears as a radially narrow circular arc with its peak surface brightness $\sim70\%$ higher than the azimuthally averaged background at the same radius (Figs.~\ref{fig:image} and \ref{fig:azimuthal}). 
In inclined disks (Fig.~\ref{fig:ipa}), we define the contrast of the vortex by normalizing the images of vortex disk models with images of disks that are axisymmetric in density distribution (Fig.~\ref{fig:azi_ave}), as even the latter have azimuthal variations caused by a finite inclination.  We find at most viewing angles the vortex peak is $50\%-70\%$ brighter than the same region in the axisymmetric disk model ($2^{\rm nd}$ and $4^{\rm th}$ columns in Fig.~\ref{fig:ipa}). The vortex is the weakest when it is on the near side and in a highly inclined disk. We note that the vortex in our model is at its peak, and its contrast is expected to be weaker at other times. 

We identify two interesting features. First, a vortex casts a shadow in the outer disk (rightmost panel in Fig.~\ref{fig:image}). 
Secondly, a vortex in an inclined disk may mimic a one-armed spiral (Fig.~\ref{fig:ipa}). Both features have been seen in scattered light observations of the HD 34282 disk (Fig.~\ref{fig:hd34282}), which has a one-armed spiral with a shadowed region on the outside \citep{deboer21, quiroz21}. The HD 34282 disk also harbors a mm continuum emission clump at roughly the same location as the one-armed spiral \citep[left panel in Fig.~\ref{fig:hd34282};][]{vanderplas17}. All evidence support that the feature may be a vortex. Vortex-like features in scattered light observations have been identified in other disks as well. For example, the feature at $r\sim0.2\arcsec$ and position angle $\sim135^\circ$ in HD 143006 disk \citep{benisty18} appears similar to our vortex at a low inclination.

We implement a few simplifications that can be improved in future works to make models more realistic. In addition to deadzone edges, vortices may also form at the edges of planet-opened gaps \citep{hammer21}. While the basic physics in the excitation of the RWI is similar in both cases \citep{ono16}, planets may deplete the inner disk, resulting in better illumination of the gap edge and the vortex by the star \citep{dong15gap}. We expect the azimuthal profiles and contrasts of the vortex in our models to remain robust, while the exact visibility of vortices in the case of planet-opened gaps needs to be further quantified. Also, while the vortex structure is similar in 2D and in 3D, in the latter case vortices may be subject to and destroyed by the elliptical instability \citep[the instability growth rate tends to be low for elongated vortices as in our models]{lesur09}. Finally, our models have low masses, and disk self-gravity is ignored. For massive disks, self-gravity may modify the structure of vortices \citep[e.g.,][]{lin11}. 

As we argue in \S\ref{sec:intro}, a promising way to test whether azimuthal asymmetries observed in disks in mm continuum emission are vortices is to compare the observed morphology of putative vortices in NIR scattered light to simulations.
While we defer modeling individual objects to a future work, we sketch a possible path forward.
For a specific system, the structure of the gas vortex candidate may be constrained by analysing gas observations \citep[e.g.,][]{muto15}, or by comparing models and observations in mm continuum emission \citep[e.g.,][]{lyra13, zhu14stone}.
Next, simulations can be carried out to produce the corresponding vortex morphology in scattered light. For relatively face-on disks with azimuthal asymmetries such as SAO 206462 \citep[e.g.,][]{pinilla15trapping, vandermarel16sao206462} and MWC 758 \citep[e.g.,][]{isella10,marino15mwc758, boehler18, casassus19}, comparing models to observations is straightforward in principle. However, observations often reveal additional features, such as spirals and shadows \citep[e.g.,][]{muto12, grady13, garufi13, benisty15, stolker16sao206462}, which need to be ``subtracted'' to facilitate the characterization of the vortex. In disks at higher inclinations, azimuthally averaged gas profiles may be obtained from gas observations to enable the production of axisymmetric disk models equivalent to Fig.~\ref{fig:azi_ave}. By comparing such models with observations, normalized scattered light maps similar to the ones in the $2^{\rm nd}$ and $4^{\rm th}$ columns in Fig.~\ref{fig:ipa} can be produced to examine the contrast of the features at the locations of the vortex candidates.


\section*{Acknowledgments}

We thank Xue-Ning Bai, Logan Francis, Pinghui Huang, Min-kai Lin, Brodie Norfolk, Bin Ren, Jessica Speedie, and Gerrit van der Plas for useful discussions and help. Computations were performed on the supercomputers provided by ComputeCanada. R.D. is supported by the Natural Sciences and Engineering Research Council of Canada and the Alfred P. Sloan Foundation. This paper makes use of data from ALMA programme 2013.1.00658.S. ALMA is a partnership of ESO (representing its member states), NSF (USA) and NINS (Japan), together with NRC (Canada) and NSC and ASIAA (Taiwan), in cooperation with the Republic of Chile. The Joint ALMA Observatory is operated by ESO, AUI/NRAO and NAOJ. The National Radio Astronomy Observatory is a facility of the National Science Foundation operated under cooperative agreement by Associated Universities, Inc.


\end{document}